\documentclass[aps,nofootinbib,preprint]{revtex4}

\usepackage{array}
\usepackage{amsmath}
\usepackage{graphicx}
\usepackage{amssymb}

\usepackage{dcolumn}
\usepackage{bm}

\usepackage{graphicx}

\begin{document}

\vspace{10cm}

\preprint{ANL-HEP-PR-09-109, NU-HEP-TH/09-14}

\title{Characteristics and Estimates of Double Parton Scattering at the Large Hadron Collider}

\author{\vspace{0.5cm} Edmond~L.~Berger}
\email{berger@anl.gov}
\affiliation{High Energy Physics Division, Argonne National Laboratory, Argonne,
IL 60439}
\author{C.~B.~Jackson}
\email{cb.jackson@mac.com}
\affiliation{High Energy Physics Division, Argonne National Laboratory, Argonne,
IL 60439}
\author{Gabe Shaughnessy}
\email{g-shaughnessy@northwestern.edu}
\affiliation{High Energy Physics Division, Argonne National Laboratory, Argonne,
IL 60439}
\affiliation{Department of Physics \& Astronomy, Northwestern University, Evanston, IL 60208}

\date{\today}

\vspace{1cm}

\begin{abstract}
We evaluate the kinematic distributions in phase space of 4-parton final-state 
subprocesses produced by double parton scattering, and we contrast these with 
the final-state distributions that originate from conventional single parton scattering.  
Our goal is to establish the distinct topologies of events that arise from these two 
sources and to provide a methodology for experimental determination of the 
relative magnitude of the double parton and single parton contributions at Large 
Hadron Collider energies.  We examine two cases in detail, the 
$b~\bar{b}~\rm{jet~ jet}$ and the 4 jet final states.   After full parton-level simulations, 
we identify a few variables that separate the two contributions remarkably well, and 
we suggest their use experimentally for an empirical measurement of the relative 
cross section.  We show that the double parton contribution falls off significantly 
more rapidly with the transverse momentum $p_T^{j1}$ of the leading jet, but, 
up to issues of the relative normalization, may be dominant  at modest values of 
$p_T^{j1}$ .    
\end{abstract}

\maketitle

%
%
\section{Introduction} 
\label{sec:introduction}
\begin{figure}[b]
\includegraphics[scale=0.6]{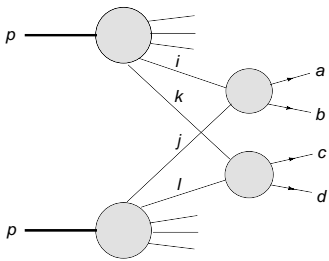}
\caption{Sketch of a double-parton process in which the active partons are 
$i$ and $k$ from one proton and $j$ and $l$ from the second proton.  The 
two hard scattering subprocess are $A(i~j \rightarrow a~b)$ and $B(k~l \rightarrow c~d)$. 
\label{fig:feyn-diag}}
\end{figure}

Double parton scattering (DPS) means that two short-distance subprocesses occur in a given hadronic interaction, with two initial partons being active from each of the incident protons in a collision at the Large Hadron Collider (LHC).  The concept is shown for illustrative purposes in Fig.~\ref{fig:feyn-diag}, 
and it may be contrasted with conventional single parton scattering (SPS) in which one short-distance subprocess occurs, with one parton active from each initial hadron.  Since the probability of single parton scattering is itself small, it is often expected that the chances are considerably suppressed for 
two or more short-distance interactions in a given collision.  However, expectations such as these bear quantitative re-examination at  the LHC where the high overall center-of-mass energy provides access to very small values of the fractional momentum $x$ carried by partons, a region in which parton densities grow rapidly.  A large contribution from double parton scattering could result in a larger than otherwise anticipated rate for multi-jet production and produce significant backgrounds in searches for signals of new phenomena.  The high energy of the LHC also provides an increased dynamic range of available phase space for detailed investigations of DPS.  

Investigations of double parton scattering have a long history 
theoretically~\cite{Goebel:1979mi, Paver:1982yp, Humpert:1983pw, Mekhfi:1983az, Humpert:1984ay, Ametller:1985tp, Halzen:1986ue, Mangano:1988sq, Godbole:1989ti, Drees:1996rw, Eboli:1997sv, 
Yuan:1997tr, Calucci:1997uw, DelFabbro:1999tf, Kulesza:1999zh, Korotkikh:2004bz, Cattaruzza:2005nu, Hussein:2006xr, Maina:2009sj, Domdey:2009bg, d'Enterria:2009hd, Gaunt:2009re}, and there is evidence for their presence in collider data from the CERN Intersecting Storage Rings~\cite{Akesson:1986iv} and Fermilab Tevatron~\cite{Abe:1997xk, D0:2009}.  A significantly greater role for double-parton processes may be expected at the LHC where higher luminosities are anticipated along with the higher collision energies.   Of substantial importance is to know empirically how large the double parton contribution may be and its dependence on relevant kinematic variables.  

Our aim is to calculate characteristic final states at LHC energies in which it may be straightforward to discern a double parton signal.  We show in this paper that double parton scattering produces an enhancement of events in regions of phase space in which the ``background'' from single parton scattering is relatively small.  If such enhancements are observed experimentally, with the kinematic dependence we predict, then we will have a direct empirical means to measure the size of the double parton contribution.  In addition to its role in general LHC phenomenology, this measurement will have an impact on the development of partonic models of hadrons, since the effective cross section for double parton scattering measures the size in impact parameter space of the incident hadron's partonic hard core.  

From the perspective of sensible rates and experimental tagging, a good process to examine should be the 4 parton final state in which there are $2$ hadronic jets plus a $b$ quark and a $\bar{b}$ antiquark, {\em viz.} $b~\bar{b}~j_1~j_2$.  If the final state arises from double parton scattering, then it is plausible that one subprocess produces the $b~\bar{b}$ system and another subprocess produces the two jets.  There are, of course, many single parton scattering (2 to 4 parton) subprocesses that can result in the $b~\bar{b}~j_1~j_2$ final state, and we look for kinematic distributions that show notable separations of the two contributions.  As we show, the correlations in the final state are predicted to be quite different between the double parton and the single parton subprocesses.  For example, the plane in which the $b~\bar{b}$ pair resides is uncorrelated with the $j_1~j_2$ plane in double parton scattering, but not in the single parton case.  

The state-of-the-art of calculations of single parton scattering is well developed whereas the phenomenology of double parton scattering is as yet much less advanced.   In the remainder of this Introduction, we first describe the approach we adopt for the calculation of double parton scattering, specializing to the proton-proton situation of the LHC.  Then we outline the paper and summarize our main results.  Our calculations are done at leading-order in perturbative QCD, adequate for the points we are trying to make.  

Making the usual factorization assumption, we express the single-parton hard-scattering differential cross section for $p~p \rightarrow a~b~X$ as  
\begin{eqnarray}
\label{eq:singlescat}
d\sigma^{SPS} = \sum_{i,j}\int f^i_p(x_1, \mu)f^j_p(x_1', \mu)
d\hat{\sigma}_{(ij \rightarrow ab)} (x_1,x_1',\mu) dx_1dx_1' .
\end{eqnarray}
Indices $i$ and $j$ run over the different parton species in each of the incident protons.  The parton-level subprocess cross sections 
$d\hat{\sigma}_{(ij \rightarrow ab)} (x_1,x_1',\mu)$ are functions of the fractional partonic longitudinal momenta $x_1$ and $x_1'$ from each 
of the incident hadrons and of the partonic factorization/renormalization scale $\mu$.   The parton distribution functions $f^i_p(x_1, \mu)$ 
express the probability that parton $i$ is found with fractional longitudinal momentum $x_1$ at scale $\mu$ in the proton; they are integrated over the intrinsic 
transverse momentum (equivalently, impact parameter) carried by the parton in the parent hadron.  

A formal theoretical treatment of double parton scattering would begin with a discussion of the hadronic 
matrix element of four field operators and an explicit operator definition of two-parton correlation functions.   This procedure would lead to a decomposition of the hadronic matrix element into non-perturbative two-parton distribution functions and the corresponding hard partonic cross sections 
for $\hat {\sigma}(i j k l \rightarrow a b c d)$.  An operator definition of two-parton correlation functions may be found in Ref.~\cite{Mueller:1985wy} where the two-parton correlation function is reduced to a product of single parton distributions.   An explicit operator definition of two-parton distributions with different values of the two fractional momenta $x_1$ and $x_2$ is presented in Ref.~\cite{Guo:1997it}, along with a model for the two-parton distributions in terms of normal parton distributions. 
In this paper, we follow a phenomenological approach along lines similar to Refs.~\cite{Goebel:1979mi, Paver:1982yp, Humpert:1983pw, Mekhfi:1983az, Humpert:1984ay, Ametller:1985tp, Halzen:1986ue, Mangano:1988sq, Godbole:1989ti, Drees:1996rw, Eboli:1997sv, 
Yuan:1997tr, Calucci:1997uw, DelFabbro:1999tf, Kulesza:1999zh, Korotkikh:2004bz, Cattaruzza:2005nu, Hussein:2006xr, Maina:2009sj, Domdey:2009bg, d'Enterria:2009hd, Gaunt:2009re}.   

In a double parton process, partons $i$ and $k$ are both active in a given incident proton.  We require the joint probability that parton $k$ carries fractional momentum $x_2$, given that parton $i$ carries fractional momentum $x_1$.  In general, this joint probability $H^{i,k}(x_1, x_2, \mu_A, \mu_B)$ should also depend on the intrinsic transverse momenta $k_{T,i}$ and $k_{T,k}$ of the two partons (or, equivalently, their impact parameters).   The hard scales $\mu_A$ and $\mu_B$ are characteristic of the two hard subprocesses in which partons $i$ and $k$ participate.  In the sections below, we discuss the choice we make of the hard-scale and do not explore in this paper theoretical uncertainties associated with higher-order perturbative contributions.  

In contrast to single parton distributions functions $f^i_p(x_1, \mu)$ for which global analyses have produced detailed information, very little is known phenomenologically about the magnitude and functional dependences of joint probabilities $H^{i,k}(x_1, x_2, \mu_A, \mu_B)$.  A common assumption made in estimates of double parton rates is to ignore possibly strong correlations in longitudinal momentum and to use the approximation 
\begin{eqnarray}
\label{eq:approx1}
H^{i,k}_p(x_1, x_2, \mu_A, \mu_B) = f^i_p(x_1, \mu_A)f^k_p(x_2, \mu_B).  
\end{eqnarray}
For reasons of energy-momentum conservation, if not dynamics, the simple factorized form of 
Eq.~(\ref{eq:approx1}) cannot be true for all values of the fractional momenta $x$.  The values of $x_2$ available to the second interaction are always limited by the values of $x_1$ in the initial interaction since $x_1 + x_2 \le 1$.  
The approximation certainly fails even at the kinematic level if both partons carry a substantial fraction of the momentum of the parent hadron.   However, it may be adequate for applications in which the values of $x_1$ and $x_2$ are small.   We remark that the momentum integral  
\begin{eqnarray}
\label{eq:sumrule}
\sum_{i,k}\int  x_1 x_2 H^{i,k}_p(x_1, x_2, \mu_A, \mu_B) dx_1dx_2  = 1 ,
\end{eqnarray}
as long as we can run the upper limits of the $x_1$ and $x_2$  integrations to $1$, independently.  The large phase space at the LHC may make it possible to explore dynamic correlations that break Eq.~(\ref{eq:approx1}).  

\begin{figure}[t]
\begin{center}
\includegraphics[scale=0.5]{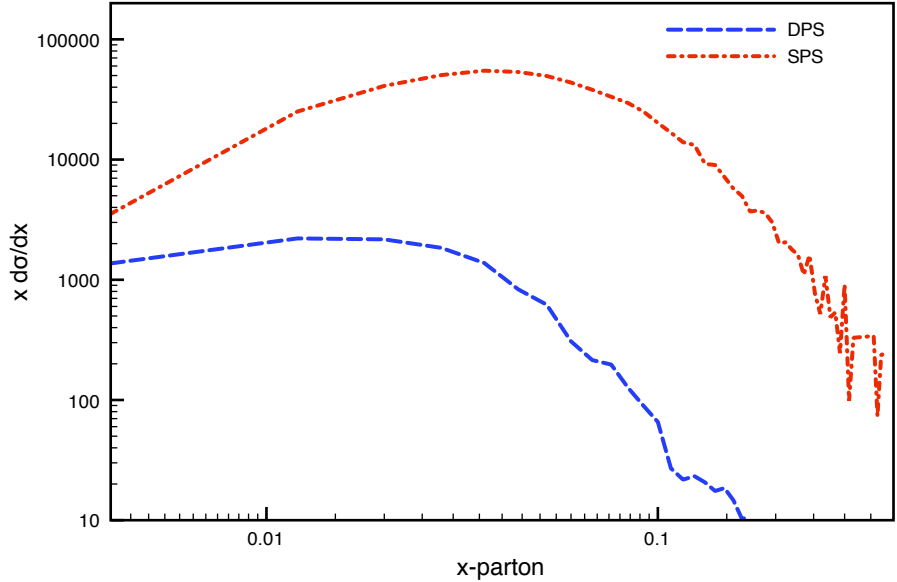}
\end{center}
\caption[]{Values of the parton longitudinal momentum fractions $x$ in the DPS and SPS events.  Most DPS 
events have low $x$ values.  The events used for this plot include the requirements $n_{\rm jet} = 4$, $n_{\rm btag} = 2$, and the threshold cuts discussed in Sec.~II.}
\label{fig:partonx}
\end{figure}

In Fig.~\ref{fig:partonx}, for the region of phase space of interest to us, we show the contributions to the $b\bar{b}jj$ cross section as a function of $x$ from both DPS and SPS, after minimal acceptance cuts are imposed (Sec.~\ref{sect:calc}).  The center-of-mass energy is $\sqrt{s} = 10$ TeV.  It is evident that the majority of DPS events are associated with low $x$ values, in essence never exceeding $0.2$.  The momentum carried off by the beam remnant is $(1 - x_1 - x_2)$ in DPS and $(1 - x)$ in SPS.  The results in Fig.~\ref{fig:partonx} show that this remnant momentum is not too different in DPS and SPS.  Thus, the use of Eq.~(\ref{eq:approx1}) in calculations of event rates at the LHC appears adequate as a good first approximation.  
While available Tevatron data on double parton scattering~\cite{Abe:1997xk, D0:2009}  are insensitive to possible correlations in $x$, the greater dynamic range at the LHC may make it possible to observe 
them.~\footnote {As emphasized in Refs.~\cite{Korotkikh:2004bz, Gaunt:2009re}, even if the approximation in Eq. (\ref{eq:approx1}) holds at one hard scale, evolution of the parton densities with 
$\mu$ will induce violations at larger scales.}  

Assuming next that the two subprocesses $A(i~j \rightarrow a~b)$ and $B(k~l \rightarrow c~d)$ are dynamically uncorrelated, we express the double parton scattering differential cross section as: 
\begin{eqnarray}
\label{eq:doubpartcross}
d\sigma^{DPS} = \dfrac{m}{2 \sigma_{\rm eff}}\sum_{i,j,k,l}\int H^{ik}_p(x_1,x_2,\mu_A,\mu_B) H_p^{jl}(x_1',x_2',\mu_A,\mu_B)
 \\ \nonumber
\times d\hat{\sigma}^A_{ij}(x_1,x_1',\mu_A) d\hat{\sigma}^B_{kl}(x_2,x_2',\mu_B) dx_1dx_2dx_1'dx_2' .
\end{eqnarray}
The symmetry factor $m$ is $1$ if the two hard-scattering subprocesses are identical 
and is $2$ otherwise.  In the denominator, there is a factor  $ \sigma_{\rm eff}$ with the 
dimensions of a cross section.  Given that one hard-scatter has taken place, $\sigma_{\rm eff}$ measures 
the size of the partonic core in which the flux of accompanying short-distance partons is confined.  It 
should be at most proportional to the transverse size of a proton.   For the first process of interest in this 
paper, $p p \rightarrow b \bar{b} j_1 j_2$, Eq.~(\ref{eq:doubpartcross}) reduces to 
\begin{eqnarray}
\label{bbjj}
d\sigma^{DPS}(p p \rightarrow b \bar{b} j_1 j_2 X) = \dfrac{d\sigma^{SPS}(p p \rightarrow b \bar{b} X) d\sigma^{SPS} (p p \rightarrow j_1 j_2 X)}{ \sigma_{\rm eff}}.  
\end{eqnarray}
Tevatron collider data~\cite{Abe:1997xk, D0:2009} yield values in the range $\sigma_{\rm eff} \sim 12$~mb.  We use this value for the estimates we make, but we emphasize that the goal should be to make an empirical determination of its value at LHC energies.  

In Sec.~\ref{sect:calc}, we present our calculation of the double parton and the single parton contributions to $p~p \rightarrow b~\bar{b}~j_1~j_2~X$.  We identify variables that discriminate the two contributions quite well.  In Sec.~\ref{sect:4jets}, we treat  the double parton and the single parton contributions to $4$ jet production, again finding that good separation is possible despite the 
combinatorial uncertainty in the pairing of jets.   We show in both cases that the double parton contribution falls off significantly more rapidly with $p_T^{j1} $, the transverse momentum of the leading jet.  For the value of $\sigma_{\rm eff} \sim 12$~mb and the cuts that we use, we find that, in the region in which it is most identifiable, double parton scattering is dominant for $p_T^{j1} < 30$~GeV in $b~\bar{b}~j_1~j_2$ at LHC energies, and $p_T^{j1} < 50$~GeV in $4$~jet production.  Our conclusions are found in Sec.~\ref{sect:conclusions}.

%
%
\section{Heavy quark pair and jet pair production in QCD.} 
\label{sect:calc}

In this section, we describe the calculation of the DPS and SPS event rates for $b\bar{b}jj$ production at the LHC.  For our purposes, light jets (denoted by $j$) are assumed to originate only from gluons or one of the four lighter quarks ($u, d, s$ or $c$) and, as stated above, we perform all calculations for the LHC with a center-of-mass energy of $\sqrt{s} = 10$ TeV.  Event rates are quoted for 10~pb$^{-1}$ of data.

\subsection{Outline of the method}

The prediction for the DPS event rate is based on the assumption that the two partonic interactions which produce the $b\bar{b}$ and $jj$ systems occur independently (as expressed in Eq.~(\ref{eq:doubpartcross})).  At leading order, the only contribution is:
\begin{equation}
(i j \rightarrow b\bar{b}) \otimes (k l \rightarrow jj)
\label{eq:DPS-LOprocess}
\end{equation} 
where the symbol $\otimes$ denotes the combination of one event each from the $b\bar{b}$ and the $jj$ final states.  In an attempt to model some of the effects expected from initial- and final-state radiation, we also account for the possibility of an additional jet which is undetected because it is either too soft 
or outside of the accepted rapidity range.  Thus, we include several other contributions to the DPS 
event:
\begin{eqnarray}
&& b\bar{b}(j) \otimes jj \,\,\,,\,\,\,  b\bar{b}j \otimes (j)j \,\,\,,\,\,\,  b\bar{b}j \otimes j(j) \\
&& b\bar{b} \otimes (j)jj \,\,\,,\,\,\,  b\bar{b} \otimes j(j)j \,\,\,,\,\,\,  b\bar{b} \otimes jj(j) \,,
\label{eq:DPS-NLOprocesses}
\end{eqnarray}
where the parentheses surrounding a jet indicate that it is undetected.  We compute processes such as 
$j j (j)$ and $b \bar{b} (j)$ at LO as 3 parton final-state processes.   

The 2 to 3 parton amplitudes for $b \bar{b} (j)$ [and $j j (j)$] diverge as the undetected jet $(j)$ becomes soft or collinear to one of the other final state partons or to an initial parton.  The divergences are removed in a full next-to-leading order (NLO) treatment, in which real emission and virtual (loop) contributions are incorporated, and the finite $b \bar{b}$, $b \bar{b} (j)$, and $b \bar{b} j$ contributions are present with proper relative normalization.  In the LO parton level simulations done in this paper, we employ a cut at the generator level to remove the divergences.   All the final state objects in the processes listed above are required to have transverse momentum $p_T  \ge 20$~GeV.  In this fashion, we model some aspects of the expected momentum imbalance between the $b$ and $\bar{b}$ arising from the 2 to 3 process $i j \rightarrow b \bar{b} j$, but we cannot claim to include the relative normalization between the $b \bar{b}$ and $b \bar{b} j$ contributions that would result from a full NLO treatment.  We leave a complete NLO analysis for future work.  

The SPS cross section is computed according to Eq.~(\ref{eq:singlescat}).  It receives contributions at lowest order from the 2 parton to 4 jet final state process:
\begin{equation} 
i j \rightarrow b\bar{b}jj \,,
\label{eq:SPS-LOprocess}
\end{equation}
and, in the case where a jet is undetected, from the 5-jet final states (computed at LO):
\begin{equation}
b\bar{b}(j)jj \,\,\,,\,\,\, b\bar{b}j(j)j \,\,\,,\,\,\, b\bar{b}jj(j) \,.
\label{eq:SPS-NLOprocesses}
\end{equation}
We also investigate the possibility of $jjjj$ and $jjjj(j)$ final state contributions to the SPS 
cross section where two of the jets ``fake'' $b$ jets.  We find that the effects from these final states are subdominant compared to the processes listed in Eqs.~(\ref{eq:SPS-LOprocess}) and (\ref{eq:SPS-NLOprocesses}).

In our numerical analysis, we use the leading-order CTEQ6L1 parton distribution functions 
(PDFs)~\cite{Pumplin:2002vw} to compute both DPS and SPS cross sections, and we evaluate all cross sections using one-loop evolution of $\alpha_s(\mu)$.  For the renormalization and factorization scales, we choose the dynamic scale:
\begin{equation}
\mu^2 = \sum_i p_{T,i}^2 + m_i^2 \,,
\label{eq:mu}
\end{equation}
where $p_{T,i}$ is the transverse momentum of the $i^{th}$ jet and $m_i = 0$ ($m_i = 4.7$ GeV) for light (bottom) jets.  In the case of roughly equal values of the transverse momenta $p_{T,i}$, Eq.~(\ref{eq:mu}) yields $\mu \sim 2 p_T$  in SPS and $\mu \sim \sqrt 2 p_T$ in DPS.  At LO there is no obviously ``right'' hard scale, and the choice in  Eq.~(\ref{eq:mu}) seems as good as any other.   

The DPS events are generated as two separate sets of events with Madgraph/Madevent~\cite{Maltoni:2002qb} and then combined as described above.  For example, at leading order, we generate events separately for $pp \to b\bar{b} X$ and $pp \to jj X$, and these events are then combined as indicated 
in Eq.~(\ref{eq:DPS-LOprocess}).  To increase the speed of the simulations, the SPS events are generated with Alpgen~\cite{Mangano:2002ea} since the SPS processes of interest are hard-coded in Alpgen, which contains more compact expressions for the squared-matrix-elements than Madgraph.  

The events accepted after generation are required to have 4 jets $n_{\rm jet} = 4$ with 2 of these tagged as $b$'s $n_{\rm btag} = 2$.  At the generator level, all the final state objects in the processes listed in Eq.~(\ref{eq:DPS-LOprocess}) through Eq.~(\ref{eq:SPS-NLOprocesses}) must have transverse momentum 
$p_T  \ge 20$~GeV, as mentioned above.  Furthermore, at the analysis level, all events (DPS and SPS) are required to pass the following acceptance cuts:
\begin{eqnarray}
p_{T,j} &\ge& 25 \,\,\,\mbox{GeV}, \,\,\, |\eta_j| \le 2.5 \\
p_{T,b} &\ge& 25 \,\,\,\mbox{GeV}, \,\,\, |\eta_b| \le 2.5 \\
\Delta R_{jj} &\ge& 0.4, \,\,\, \Delta R_{bb} \ge 0.4
\label{eq:cuts}
\end{eqnarray}
where $\eta_i$ is the jet's pseudorapidity, and $\Delta R_{ij}$ is the separation in the 
azimuthal angle ($\phi$) - pseudorapidity plane between jets $i$ and $j$:
\begin{equation}
\Delta R_{ij} = \sqrt{ (\eta_i - \eta_j)^2 + (\phi_i - \phi_j)^2 } \,.
\label{eq:DeltaR}
\end{equation}
We model detector resolution effects by smearing the final state energy according to:
\begin{equation}
{\delta E \over E} = {a \over \sqrt{E/\rm{GeV}}} \oplus b,
\end{equation}
where we take $a=50\%$ and $b=3\%$ for jets.  To account for $b$ jet tagging efficiencies, we assume a $b$-tagging rate of 60\% for $b$-quarks with $p_T > 20\text{ GeV}$ and $|\eta_{b}| < 2.0$.  We also apply a mistagging rate for charm-quarks as:
\begin{equation}
\epsilon_{c\to b} =10\% \quad\quad \text{ for } p_T(c) > 50 \text{ GeV}\\
\end{equation}
while the mistagging rate for a light quark is: 
\begin{eqnarray}
\epsilon_{u,d,s,g\to b} &= 0.67\% \quad\quad \text{ for }& p_T(j) < 100 \text{ GeV} \\
\epsilon_{u,d,s,g\to b} &= 2\% \quad\quad \quad \text{ for }& p_T(j)> 250 \text{ GeV} .
\end{eqnarray}
Over the range $100\text{ GeV}<p_T(j)<250\text{ GeV}$, we linearly interpolate the fake rates given above~\cite{Baer:2007ya}.

\subsection{Properties of SPS and DPS in $b~\bar{b}~j~j$}
\label{sect:results}

Having detailed the calculation of the $b\bar{b}jj$ event rates from DPS and SPS, we now discuss some of the distinguishing characteristics of the two contributions.  First, however, it is important to check that our simulations of DPS events are not introducing an {\it artificial} correlation between the $b\bar{b}$ and $jj$ final states.  We do this by inspecting the angle $\Phi$ between the plane defined by the $b\bar{b}$ system and the plane defined by the $jj$ system.  If the two scattering processes $i j \rightarrow b \bar{b}$ and $k l \rightarrow j j$ which produce the DPS final state are truly independent, one would expect to see a flat distribution in the angle $\Phi$.  By contrast, many diagrams, including some with non-trivial spin correlations, contribute to the 2 parton to 4 parton final state in SPS, and naively one would expect some correlation between the two planes.  To avoid possible effects from boosting to the lab frame, we define the two planes in the partonic center-of-mass frame.

We specify the planes by using the three-momenta of the outgoing jets.  Then, the angle between the two planes defined by the $jj$ and $b\bar{b}$ systems is:
\begin{equation}
\cos \Phi = \hat n_3(j_1,j_2) \cdot \hat n_3(b_1,b_2),
\label{eq:cosPhi_SPS}
\end{equation}
where $\hat n_3(x,y)$ is the unit three-vector normal to the plane defined by the $x-y$ system.   

The normal is undefined when $j_1$ and $j_2$ are back-to-back or $b_1$ and $b_2$ are back-to-back, as occurs in a large fraction of the DPS events.    
Therefore, when $\cos \phi_{(x,y)} <  -0.9$, we use a different procedure.    We use the three-momentum of one of the incoming partons along with the three-momentum of one of the outgoing $b$ quarks to define the $b\bar{b}$ plane.  Let $q_b$ be the three-momentum of an incoming parton, and $p_b$ be the three-momentum of the final-state $b$ (or $\bar{b}$) quark.  We then define $\phi_{p_b, q_b}$ to be the azimuthal angle of the three-vector normal to the $q_b - p_b$ plane.  Note that we use $\phi$ here since the normal to any three-vector and the beam-line will be transverse to the beam-line (not the case in the SPS process).  In this way, the jet which is not used to define the plane is guaranteed to lie in the plane.  The plane for the $jj$ system is defined in an analogous manner.  Finally, the angle between the planes is then:
\begin{equation}
\Phi = |\phi_{p_j,q_j} - \phi_{p_b,q_b}|\,.
\label{eq:Phi_DPS}
\end{equation}

\begin{figure}[t]
\begin{center}
\includegraphics[width=0.59\textwidth]{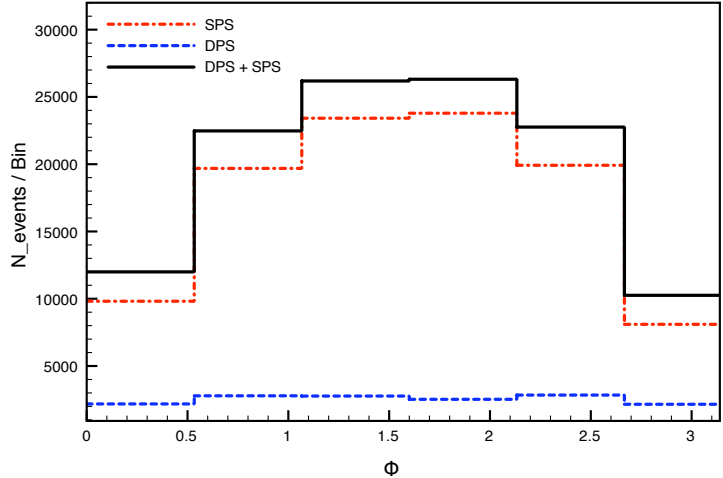}
\end{center}
\caption[]{Event rate as a function of the angle between the two planes
  defined by the $b\bar{b}$ and $jj$ systems.  In SPS events, there is a correlation among the planes which is absent for DPS events.  }
\label{fig:Phi-planes}
\end{figure}

In Fig.~\ref{fig:Phi-planes}, we display the number of events as a function of the angle 
between the two planes.  There is an evident correlation between the two planes in SPS, 
while the distribution is flat in DPS, indicative that the two planes are uncorrelated.

\begin{figure}[ht]
 \begin{center}
\includegraphics[width=0.59\textwidth]{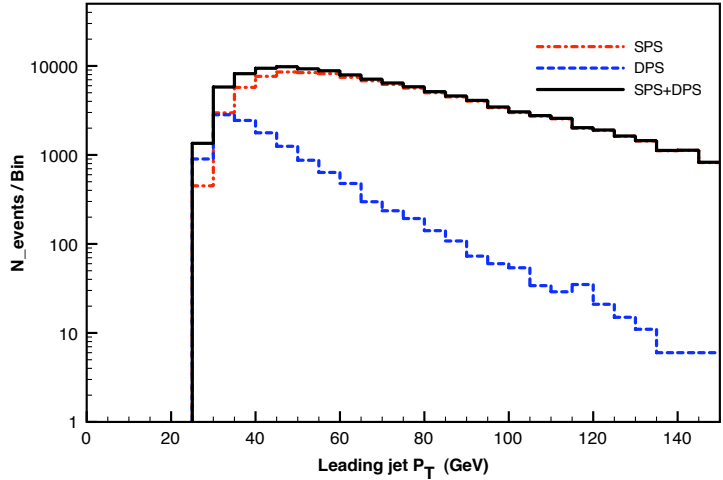}
\end{center}
\caption[]{The transverse momentum $p_T$ distribution of the leading jet in $jjb\bar b$ after minimal cuts. }
\label{fig:ptj_1}
\end{figure}

Another interesting difference between DPS and SPS is the behavior of event rates as a function of transverse momentum.  As an example of this, in Fig.~\ref{fig:ptj_1}, we show the transverse momentum distribution for the leading jet (either a $b$ or light $j$) for both DPS and SPS.  Several characteristics are evident.  First, SPS produces a relatively hard spectrum, and for the value of $\sigma_{\rm eff}$ and the cuts that we use, we see that SPS tends to dominate over the full range of transverse momentum considered.  On the other hand, DPS produces a much softer spectrum which (up to issues of normalization in the form of $\sigma_{\rm eff}$) can dominate at small values of transverse momentum.  The cross-over between the two contributions to the total event rate is $\sim 30$ GeV for the acceptance cuts considered here.   A smaller (larger) value of $\sigma_{\rm eff}$ would move the cross-over to a larger (smaller) value of the transverse momentum $p_T^{j1}$ of the leading jet.    

\subsection{Distinguishing variables}
\label{sect:variables}

\begin{figure}[htbp]
\begin{center}
\includegraphics[width=0.49\textwidth]{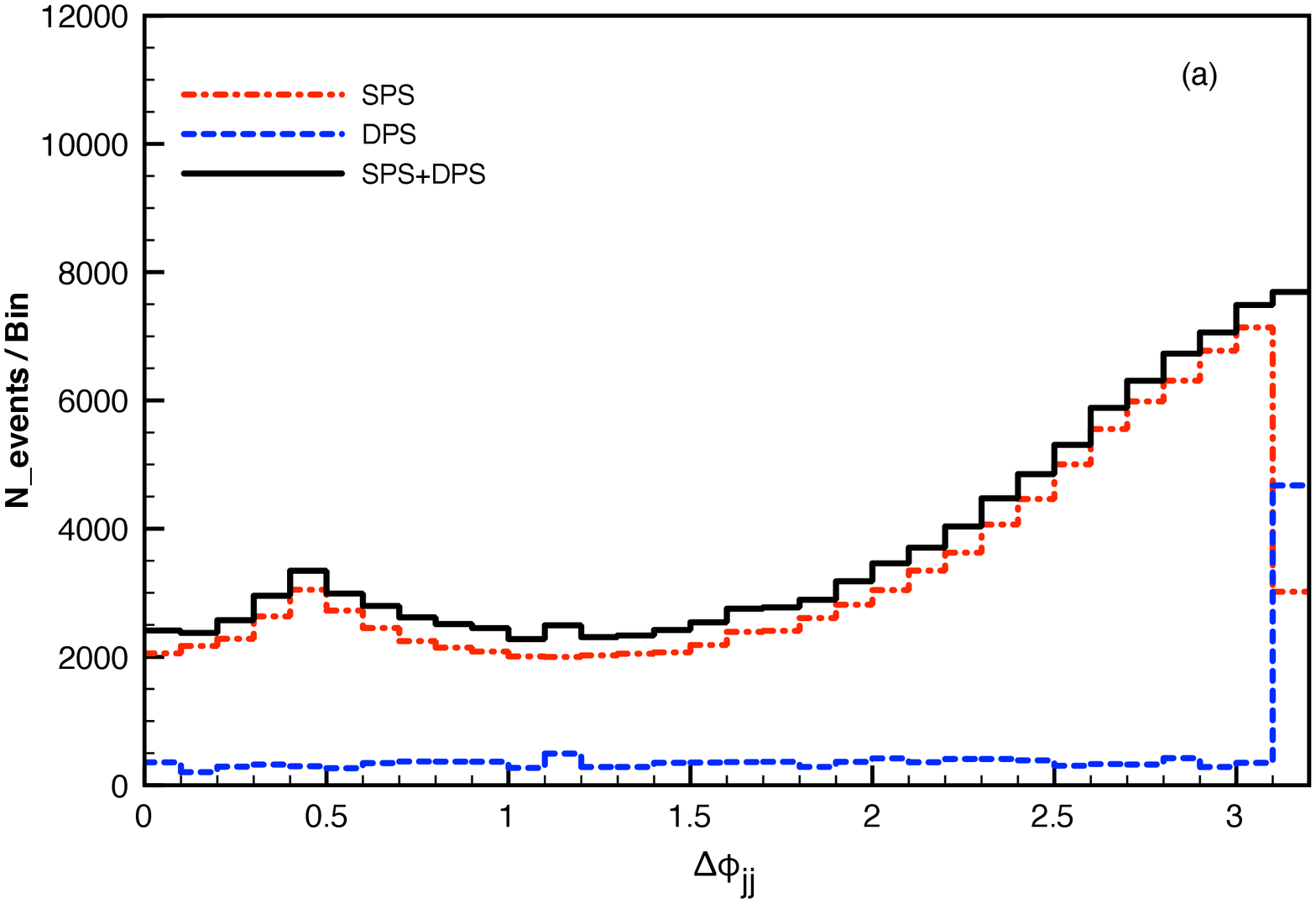}
\includegraphics[width=0.49\textwidth]{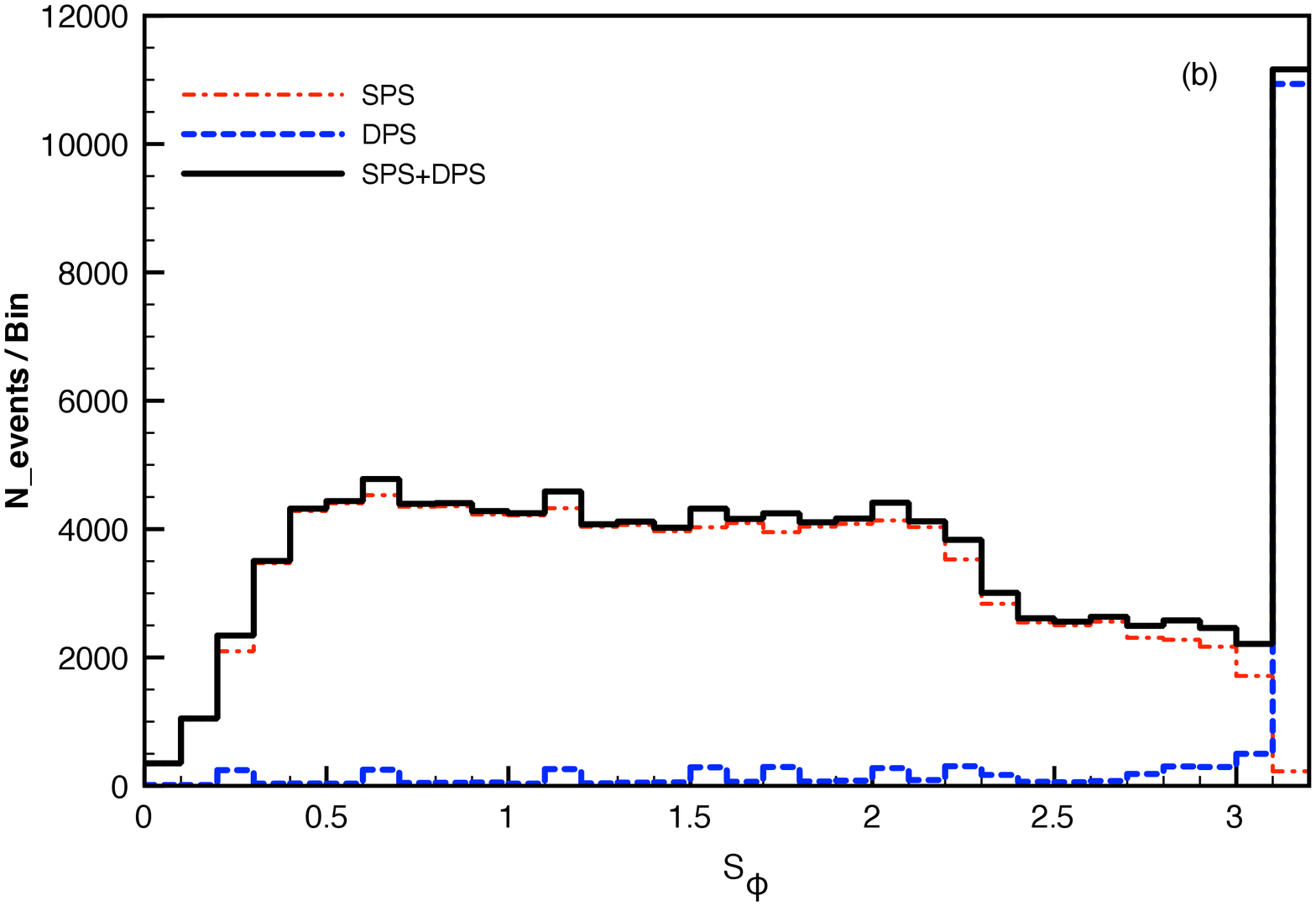}
\caption{(a) The difference $\Delta \phi$ in the azimuthal angles of light jet pairs for DPS and both SPS+DPS events.  The dijet pairs are back-to-back in DPS events.  (b) The variable $S_\phi$ for DPS and SPS+DPS events provides a stronger separation of the underlying DPS events from the total sample when compared to $\Delta\phi$ for any pair.}
\label{fig:delphi}
\end{center}
\end{figure}

We turn next to the search for variables that may allow for a clear separation of the DPS and SPS contributions.  Since the topology of the DPS events includes two $2\to2$ hard scattering events, the two pairs of jet objects are roughly back-to-back.  We expect the azimuthal angle between the pairs of jets corresponding to each hard scattering event to be strongly peaked near $\Delta \phi_{jj} \sim \Delta \phi_{bb} \sim \pi$.  Real radiation of an additional jet, where the extra jet is missed because it fails the threshold or acceptance cuts, allows smaller values of $\Delta \phi_{jj}$.  The relevant distribution is shown for light jets (non $b$-tagged) in Fig.~\ref{fig:delphi}a. There is a clear peak near $\Delta\phi_{jj}=\pi$ for DPS events, while the events are more broadly distributed in SPS events.  The secondary peak near small $\Delta\phi_{jj}$ arises from gluon splitting which typically produces nearly collinear jets.  The suppression at still lower $\Delta\phi_{jj}$ comes from the isolation cut $\Delta R_{jj} > 0.4$.  

The separation of DPS events from SPS events becomes more pronounced if information is used from both the $b\bar{b}$ and $jj$ systems.  As an example, we consider the distribution built from a combination of the azimuthal angle separations of both $jj$ and $b\bar b$ pairs, using a variable adopted from Ref.~\cite{D0:2009}:
\begin{equation}
S_{\phi}={1\over \sqrt 2} \sqrt{\Delta \phi(b_1,b_2)^2+\Delta \phi(j_1,j_2)^2}.
\end{equation}
In Fig.~\ref{fig:delphi}b, we present a distribution in $S_{\phi}$ for both DPS and SPS+DPS events.  Again, as in the case of the $\Delta \phi$ distribution, we see that the SPS events are broadly distributed across the allowed range of $S_\phi$.  However, the combined information from both the $b\bar{b}$ and $jj$ systems shows that the DPS events produce a sharp and substantial peak near $S_\phi \simeq \pi$ which is well-separated from the total sample.

The narrow peaks near $\Delta\phi_{jj}=\pi$ in Fig.~\ref{fig:delphi}a and near $S_{\phi} = 1$ in 
Fig.~\ref{fig:delphi}b will be smeared somewhat once soft QCD radiation and other higher-order terms are included in the calculation.  

Another possibility for discerning DPS is the use of the total transverse momentum of both the $b\bar{b}$ and $jj$ systems.  At lowest order for a $2 \to 2$ process, the vector sum of the transverse momenta of the final state pair vanishes.  In reality, radiation and momentum mismeasurement smear the expected peak near zero.  Nevertheless, we still expect DPS events to show a distribution in the transverse momenta of the jet pairs that is reasonably well-balanced.  To encapsulate this expectation for both light jet pairs and $b$-tagged pairs, we use the variable~\cite{D0:2009}:
\begin{equation}
S_{p_T}^\prime={1\over \sqrt 2} \sqrt{\left({|p_T(b_1,b_2)|\over |p_T(b_1)|+|p_T(b_2)|}\right)^2+\left({|p_T(j_1,j_2)|\over |p_T(j_1)|+|p_T(j_2)|}\right)^2}.
\label{eq:sptprime}
\end{equation}
Here $p_T(b_1,b_2)$ is the vector sum of the transverse momenta of the two final state $b$ jets, and $p_T(j_1,j_2)$ is the vector sum of the transverse momenta of the two (non $b$) jets.  

\begin{figure}[ht]
 \begin{center}
\includegraphics[width=0.49\textwidth]{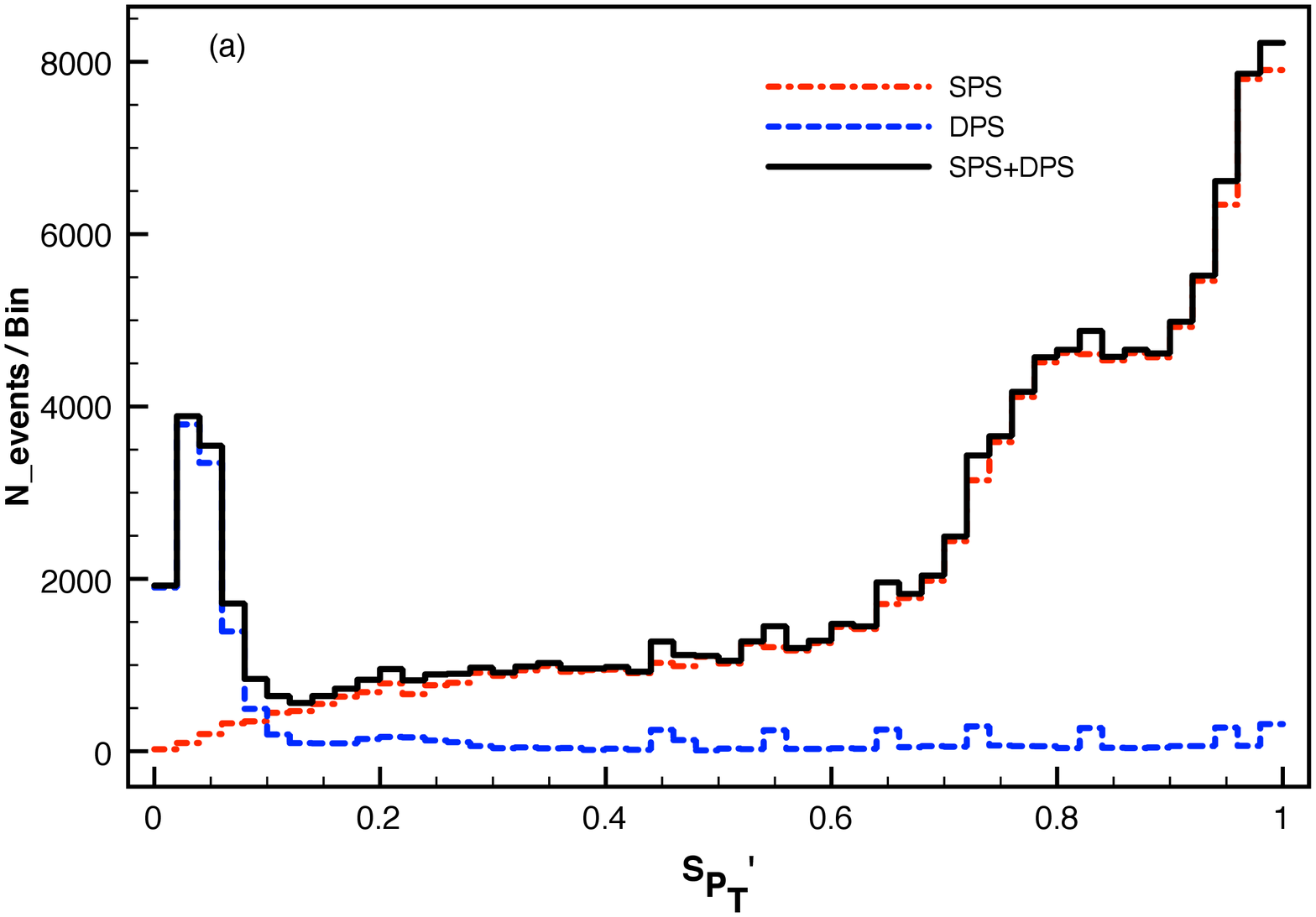}
\includegraphics[width=0.49\textwidth]{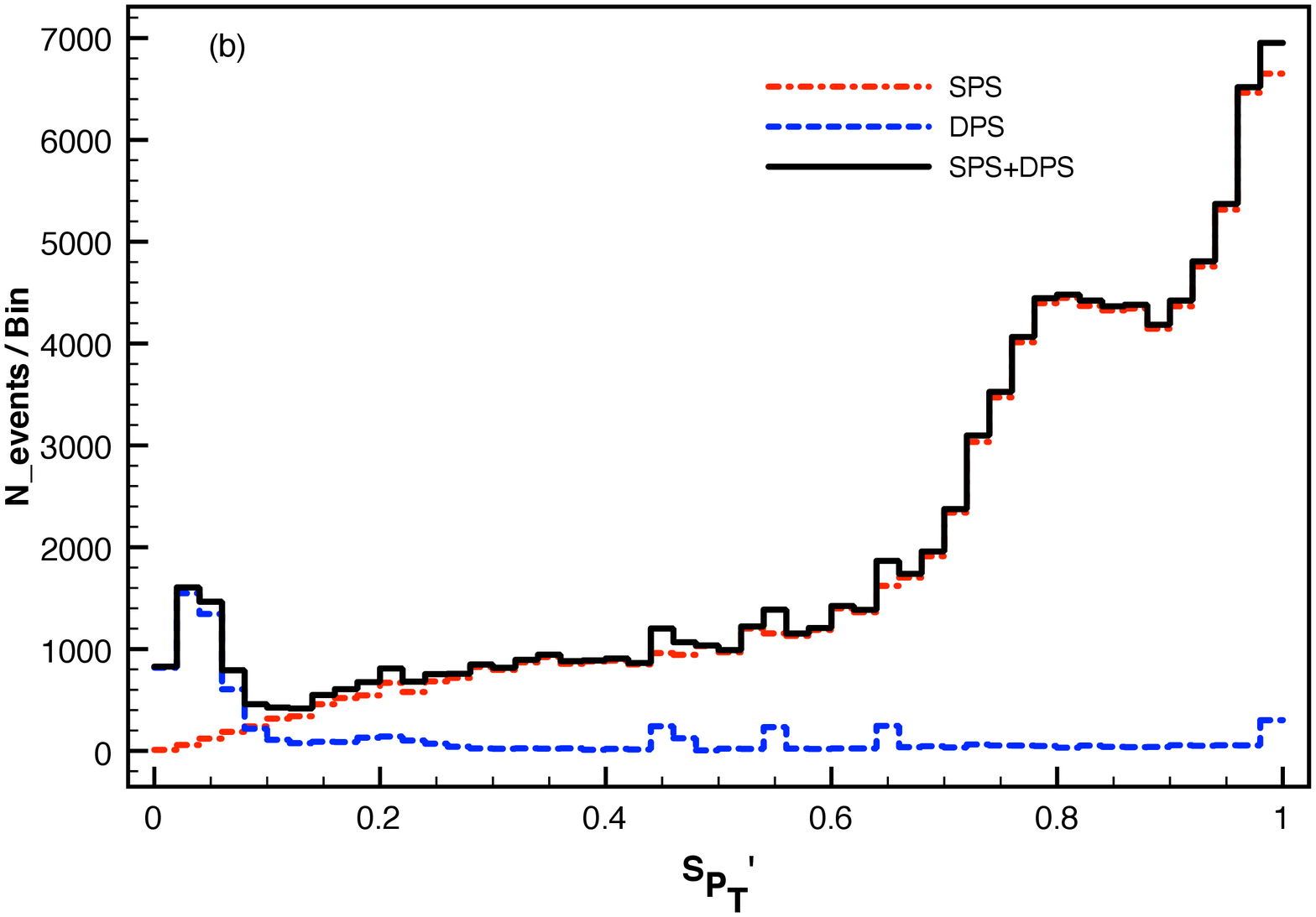}
\end{center}
\caption[]{Distribution of events in $S_{p_T}^\prime$ for the DPS and SPS samples.  Due to the back-to-back nature of the $2\to2$ events in DPS scattering, the transverse momenta of the jet pair and of the 
$b$-tagged jet pair are small, resulting in a small value of $S_{p_T}^\prime$.  In (a) we show the 
$S_{p_T}^\prime$ distribution for our standard cuts, and in (b) we increase the cut on the transverse momentum of the leading jet, $p_T^{j1} > 40$ GeV.  The fraction of DPS  events in the whole sample decreases with increasing $p_T^{j1}$.}
\label{fig:sptprimecut}
\end{figure}

The distribution in $S_{p_T}^\prime$ is shown in Fig.~\ref{fig:sptprimecut}.  As expected, we observe that the DPS events are peaked near $S_{p_T}^\prime\sim0$ and are well-separated from the total sample.  The SPS events, on the other hand, tend to be far from a back-to-back configuration and, in fact, are peaked near $S_{p_T}^\prime\sim1$.  This behavior of the SPS events is presumably related to the fact that a large number of the $b\bar{b}$ or $jj$ pairs arise from gluon splitting which yields a large $p_T$ imbalance and, thus, larger values of $S_{p_T}^\prime$.

In this subsection, we find that extraction of the DPS ``signal'' for $b\bar{b}jj$ production from the SPS ``background'' can be enhanced by combining information from both $b\bar{b}$ and $jj$ systems.  Our simulations suggest that the variable $S_{p_T}^\prime$ may be a more effective discriminator than 
$S_{\phi}$.   However, given the leading order nature of our calculations and the absence of smearing associated with initial state soft radiation, this picture may change and a variable such as $S_\phi$ (or some other variable) may become a clearer signal of DPS at the LHC.  Realistically, it would be valuable to study both distributions once LHC data are available in order to determine which is more instructive.  In the following, we use the clear separation shown in Fig.~\ref{fig:sptprimecut} in our exploration of the distinct properties of DPS and SPS events.  

\subsection{Two-dimensional distributions}
\label{sect:scatter}

The evidence in Fig.~\ref{fig:delphi} and Fig.~\ref{fig:sptprimecut} for distinct regions of DPS dominance prompts the search for greater discrimination in a plane represented by a two dimensional distribution of one variable against another.  We examined scatter plots involving the inter-plane angle $\Phi$, the jet-jet azimuthal angle difference $\Delta \phi_{jj}$,  $S_{\phi}$, and $S'_{p_T}$.  Strong kinematic correlations are evident in the plot of  $S_{\phi}$ {\em vs.} $S'_{p_T}$ at the level of our leading order calculation, and we observe no additional separation of DPS and SPS beyond that evident in 
Figs.~\ref{fig:delphi} and~\ref{fig:sptprimecut}.  Likewise, there are strong correlations between 
$\Delta \phi_{jj}$ and $S_{\phi}$.  

\begin{figure}[ht]
 \begin{center}
\includegraphics[width=0.99\textwidth]{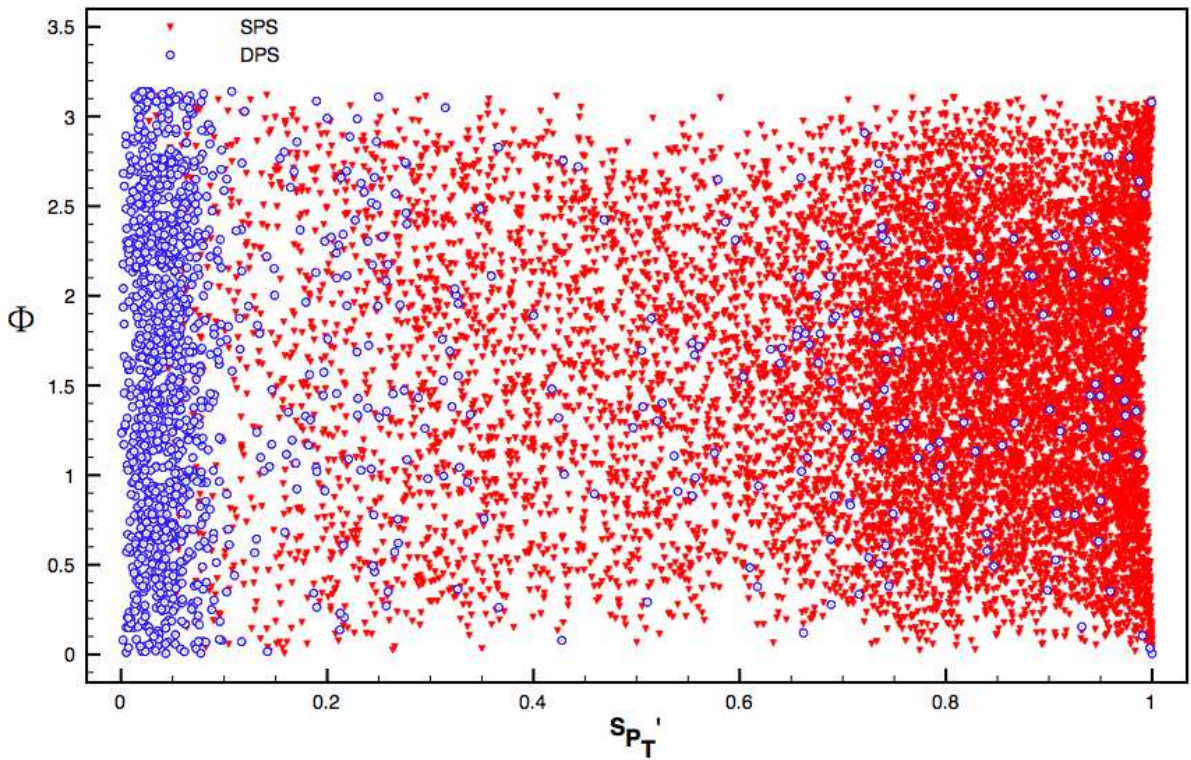}
\end{center}
\caption[]{Two-dimensional distribution of events in the variables $\Phi$ and $S_{p_T}^\prime$ for the DPS and SPS samples.}
\label{fig:scatter1}
\end{figure}

One scatter plot with interesting features is displayed in 
Fig.~\ref{fig:scatter1}.   The DPS events are seen to be clustered near $S'_{p_T} = 0$ and are uniformly distributed in $\Phi$.  The SPS events peak toward $S'_{p_T} = 1$ and show a roughly $\sin \Phi$ character.   While already evident in Figs.~\ref{fig:Phi-planes} and~\ref{fig:sptprimecut}, these two  
features are more apparent in the scatter plot Fig.~\ref{fig:scatter1}.  Moreover, the 
scatter plot shows a valley of relatively low density between $S'_{p_T} \sim 0.1$ and $\sim 0.4$. 
In an experimental one-dimensional $\Phi$ distribution such as Fig.~\ref{fig:Phi-planes}, one would see the sum of the DPS and SPS contributions.  If structure is seen in data similar to that shown in the scatter plot Fig.~\ref{fig:scatter1}, one could make a cut at $S'_{p_T} < 0.1$ or $0.2$ and verify whether the experimental distribution in $\Phi$ is flat as expected for DPS events.  
 
In Fig.~\ref{fig:ptj_1}, we show that DPS produces a softer transverse momentum distribution for the leading jet (either a $b$ or light $j$).  In data one would see only the sum of the DPS and SPS components in a plot like Fig.~\ref{fig:ptj_1}.  A scatter plot of 
$S_{p_T}^\prime$ {\em vs.} the transverse momentum of the leading jet motivates an empirical separation of the two components.  In Figs.~\ref{fig:sptprimecut}(a) and~\ref{fig:sptprimecut}(b) we compare the  $S_{p_T}^\prime$ distributions for two different selections on the transverse momentum $p_T^{j1}$ of the leading jet in the $b \bar{b} j j$ sample.  This comparison of the distributions confirms that  events in the DPS region, defined empirically by the region $S_{p_T}^\prime < 0.1$~or~$0.2$, fall off more steeply with $p_T^{j1}$ than the rest of the sample.  It will be important and interesting to see whether the selection $S_{p_T}^\prime < 0.1$~or~$0.2$ in LHC data also produces events that show a more rapid decrease with $p_T^{j1}$.  

\begin{figure}[ht]
 \begin{center}
\includegraphics[width=0.49\textwidth]{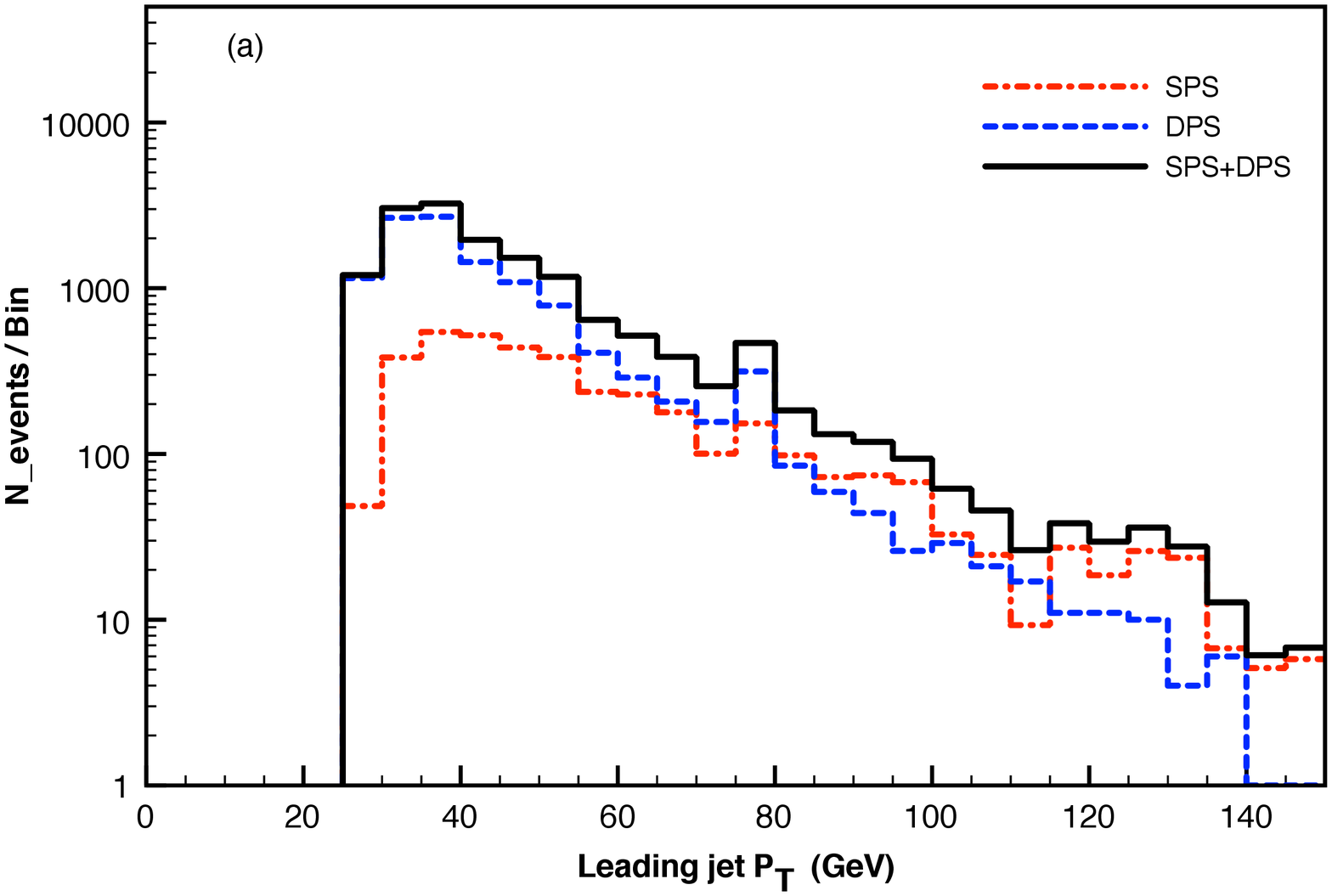}
\includegraphics[width=0.49\textwidth]{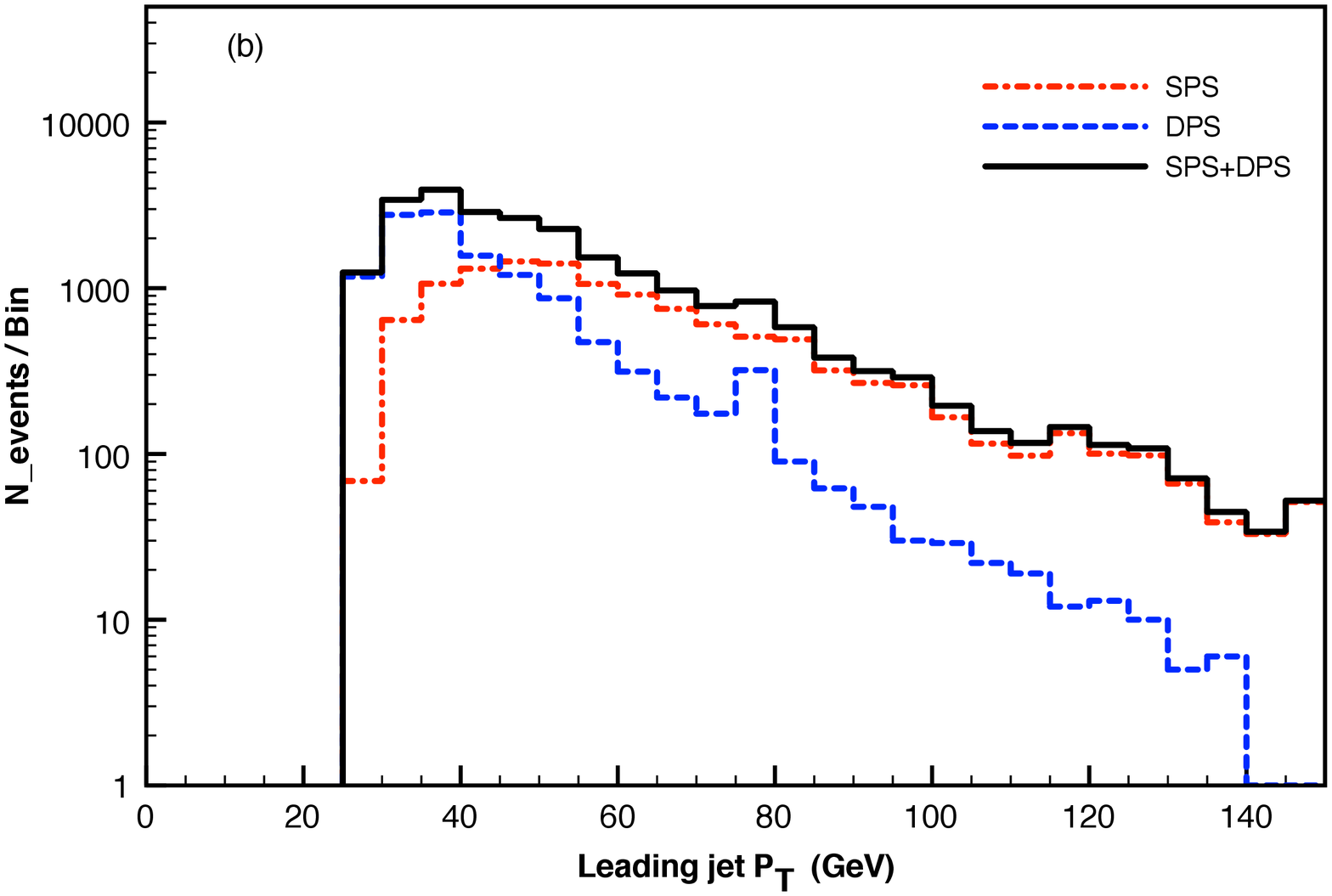}
\end{center}
\caption[]{The distribution in the transverse momentum of the leading jet $p_T^{j1}$ for (a) $S_{p_T}^\prime < 0.2$ and (b) $S_{p_T}^\prime< 0.4$.  As the signal region becomes more dominated by SPS events (i.e. moving from (a) to (b)), the resulting distribution becomes harder and shifts the SPS-DPS cross-over from $\sim80$ GeV to $\sim40$ GeV. }
\label{fig:ptjcrossover}
\end{figure}

The leading-jet transverse momentum distributions are shown in Figs.~\ref{fig:ptjcrossover}(a)~and~\ref{fig:ptjcrossover}(b) for two different cuts on  $S_{p_T}^\prime$.  In both cases, we see that the SPS sample has a broader distribution in $p_T^{j1}$ and that the DPS sample dominates for small enough values of $p_T^{j1}$.  For our chosen value of $\sigma_{\rm eff} \sim 12$~mb, and for cuts we employ, the crossover points are roughly $80$~GeV for $S_{p_T}^\prime < 0.2$ and $40$~GeV for 
$S_{p_T}^\prime < 0.4$.   

%
%
\section{Four Jet Production} 
\label{sect:4jets}

In addition to $b\bar b jj$, we can also ask how important DPS can be for a generic $4j$ final-state, where none of the jets are $b$-tagged.  In this section, we describe our calculation of the double parton scattering and the single parton scattering contributions to the production of a $4j$ final state, for which the cross section is larger.  Our exposition can be brief since we repeat the procedure described in some detail in Sec.~\ref{sect:calc}.  

\subsection{Outline of the method}

The DPS process for $4j$ production is topologically equivalent to $b \bar b jj$.  However, in the $4j$ system, we lose the $b$-tagging ability that reduces the combinatorial background in $b\bar b jj$, and the prospects for isolating and measuring DPS over the SPS background may appear less promising.  Fortunately, in going from the $b\bar b$ subprocess to the $jj$ subprocess, a much larger DPS rate is possible due to the much larger cross section for $jj$ production.  As we show below, we find that the DPS signature can be extracted in this $4j$ mode as well.

The DPS cross section for $4j$ production receives contributions from the following sub-processes at the lowest order:
\begin{equation}
jj \otimes jj  \,\,\,,\,\,\,  b\bar b \otimes jj ,
\label{eq:DPS-LOprocess-4j}
\end{equation} 
where both $b$-quarks fail the $b$-tag.  We do not include the $b\bar b \otimes b\bar b$ process due to its relatively small rate ($\sim 0.14$ nb).  This rate is further reduced by requiring no $b$-tags, yielding roughly 40 events in the 10 pb$^{-1}$ of luminosity assumed here.  

Following Sec.~\ref{sect:calc}, we account for the possibility of an additional jet which is 
undetected because it is too soft or outside of the accepted rapidity range.  Thus, we 
include several other contributions to the DPS cross section:
\begin{eqnarray}
&& jjj \otimes (j)j \,\,\,,\,\,\,  jj(j) \otimes jj \,,\\
&& b\bar b j \otimes j(j) \,\,\,,\,\,\,  b\bar b (j) \otimes jj, \\
&& b\bar b \otimes j(j)j \,\,\,,\,\,\,  b(\bar b) \otimes jjj \,\,\,,\,\,\,  (b)\bar b \otimes jjj \,.
\label{eq:DPS-NLOprocesses-4j}
\end{eqnarray}
where the parentheses surrounding a jet signify that it is not detected.

The SPS cross section receives contributions at lowest order from the final state:
\begin{equation} 
jjjj \,\,\, ,b\bar{b}jj \,,
\label{eq:SPS-LOprocess-4j}
\end{equation}
where both $b$-quarks fail the $b$-tag, and, in the case where a jet is not detected, from the 
final states:
\begin{equation}
b\bar{b}(j)jj \,\,\,,\,\,\, (b)\bar{b}jjj \,\,\,,\,\,\, b(\bar{b})jjj \,\,\,,\,\,\, (j)jjjj\,.
\label{eq:SPS-NLOprocesses-4j}
\end{equation}
We refer to Sec.~\ref{sect:calc} for the specification of acceptance cuts and detector resolution, and for our treatment of the potential divergences present in the amplitudes for the processes in 
Eqs.~(\ref{eq:DPS-LOprocess-4j})-(\ref{eq:SPS-NLOprocesses-4j}).  

\subsection{Results}

\begin{figure}[htb]
 \begin{center}
\includegraphics[width=0.59\textwidth]{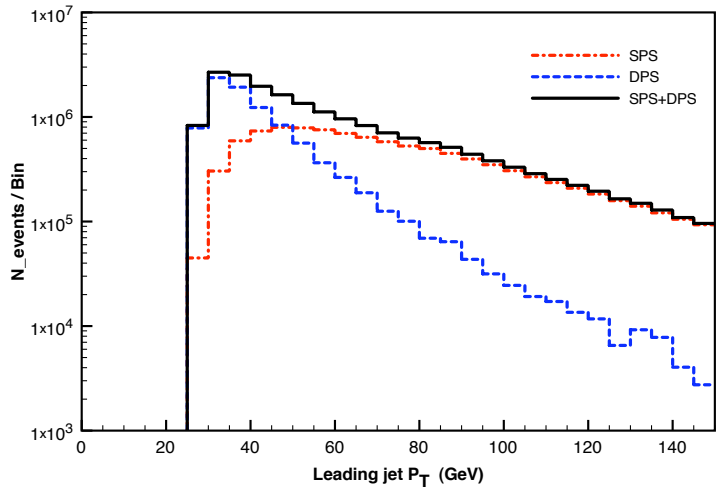}
\end{center}
\caption[]{As in Fig.~\ref{fig:ptj_1}, but for $4j$ events.  Similar to the $b \bar b j j$ sample, the SPS sample exhibits a harder $p_T$ spectrum.}
\label{fig:ptj_1-4j}
\end{figure}

Similar to the $b\bar b jj$ process, the leading jet in the $4j$ DPS sample is typically softer than in the SPS channels (see Fig.~\ref{fig:ptj_1-4j}).  In this case, again using $\sigma_{\rm eff} = 12$~mb, we find that the cross-over between DPS and SPS dominance occurs near $p_T \simeq 50$ GeV, higher than in the $b \bar{b} j j$ case shown in Fig.~\ref{fig:ptj_1}.  
Improvement in the separation between DPS and SPS in the $4j$ case can be achieved with an analogous version of the $S_{p_T}^\prime$ variable introduced in Eq.~(\ref{eq:sptprime}):
\begin{equation}
S_{p_T}^\prime={1\over \sqrt 2} \sqrt{\left({|p_T(j_a,j_b)|\over |p_T(j_a)|+|p_T(j_b)|}\right)^2+\left({|p_T(j_c,j_d)|\over |p_T(j_c)|+|p_T(j_d)|}\right)^2}.
\label{eq:sptprime2}
\end{equation}
Here $p_T(j_a,j_b)$ is the vector sum of the transverse momenta of two final state jets, $a$ and $b$, chosen among the four.  The remaining $c$ and $d$ jets are then fixed.  This choice is unique if a separation of the two hard interactions is possible.  In the $b\bar b jj$ system, the separation into the $b\bar b$ and $jj$ subsystems via $b$-tagging removed most of the degeneracy (some degeneracy still remained via tagging efficiencies or light jet mistagging).  In the $4j$ system, the degeneracy can at first glance be problematic as there are 3 possible pairings of the four jets.  

\begin{figure}[htb]
 \begin{center}
\includegraphics[width=0.59\textwidth]{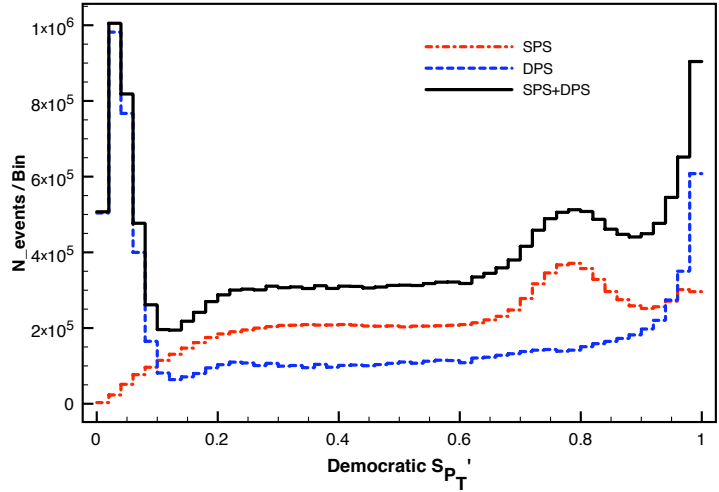}
\end{center}
\caption[]{The democratic $S_{p_T}^\prime$ distribution for $4j$ events shows much more combinatorial background than in the $b \bar b jj$ events.  Even after accepting two mis-matched jet pairs, we see that the DPS and SPS samples can still be separated well.}
\label{fig:sptprimedem}
\end{figure}
One might be tempted to take the pairing of jets which minimizes the value of $S_{p_T}^\prime$.  Unfortunately, this choice places a bias on the distribution that makes it potentially problematic to trust the discrimination.  Instead, to construct $S_{p_T}^\prime$ we take all three combinations of pairings, which includes one ``correct'' pairing and two incorrect pairings in the DPS process.  This ``democratic'' $S_{p_T}^\prime$  distribution is shown in Fig.~\ref{fig:sptprimedem} and is re-weighted by 1/3 for proper normalization.  As in the $b\bar b jj$ case, we see that the DPS distribution peaks near $S_{p_T}^\prime\sim0$, indicative that two back-to-back hard interactions are present.  In addition to this expected feature, we also see a continuum that extends above $S_{p_T}^\prime\sim 0.1$, associated with the wrong combination taken in the democratic approach.  
In Fig.~\ref{fig:sptprimedem} we see that DPS produces a secondary peak at $S_{p_T}^\prime\sim1$, 
not present in the $b \bar{b} j j$ case in Fig.~\ref{fig:sptprimecut}.  It appears to arise from the wrong pairings of jets associated with the combinatorial background.  In these instances, the wrong combination of two jets that are close together in $\Delta R$, 
meaning that their momenta are aligned, can maximize the value of $S_{p_T}^\prime$.
Overall, we see that the DPS peak near $S_{p_T}^\prime = 0$ provides a good means to separate DPS events from SPS events.

\begin{figure}[ht]
 \begin{center}
\includegraphics[width=0.49\textwidth]{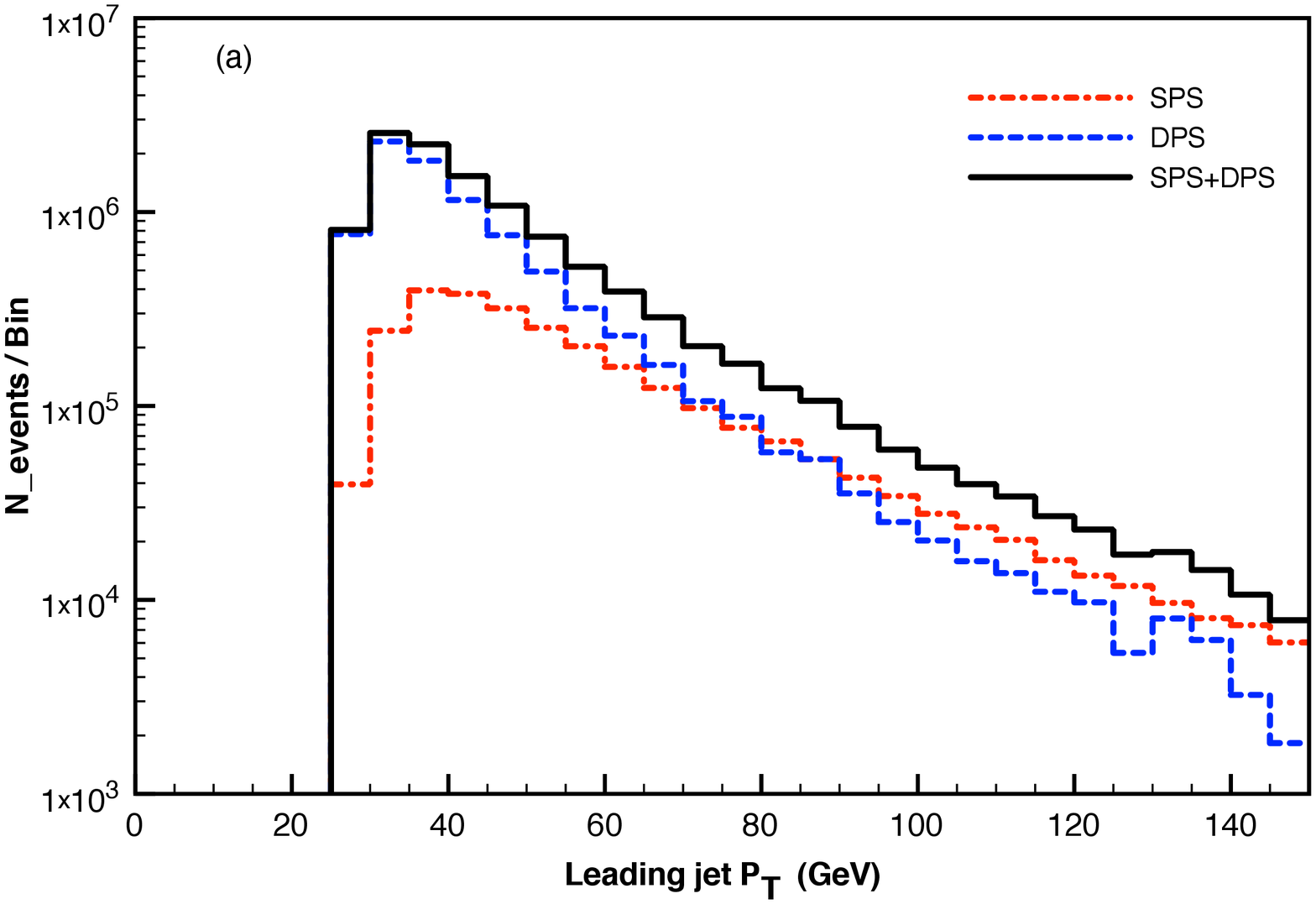}
\includegraphics[width=0.49\textwidth]{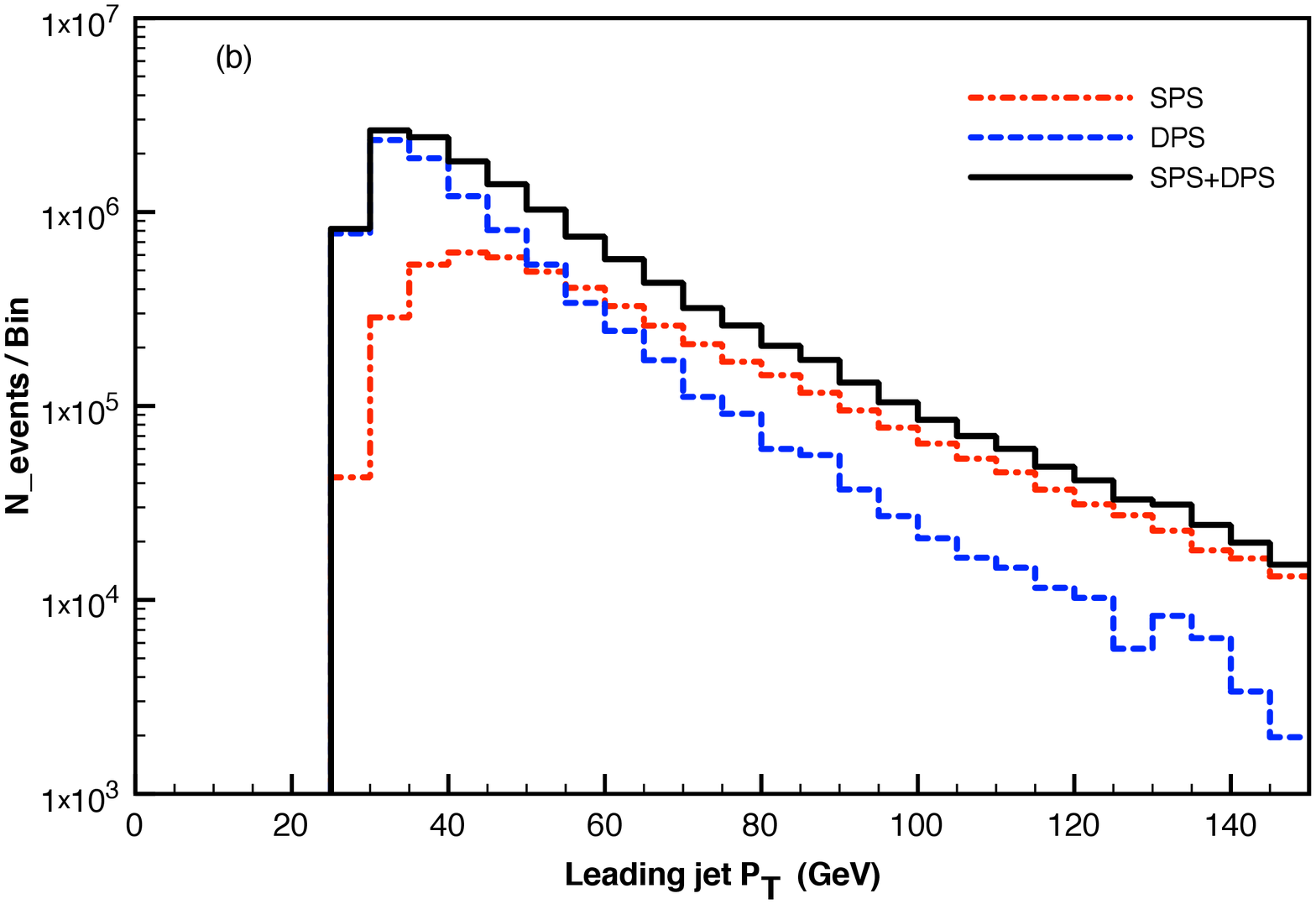}
\end{center}
\caption[]{As in Fig.~\ref{fig:ptjcrossover}, but for $4j$ events with (a) democratic $S_{p_T}^\prime < 0.2$ and (b) democratic $S_{p_T}^\prime< 0.4$.  As in $b\bar b j j$ events, as one increases the cut on $S_{p_T}^\prime$, the SPS fraction increases and the total distribution is harder. }
\label{fig:ptj_1-4j-crossover}
\end{figure}
As in the $b \bar{b} j j$ case, we inspect the distribution in the $p_T$ of the leading jet after cuts on the $S_{p_T}^\prime$ variable.  Since there are three jet pairings per event, we now require that at least one of the three pairings has $S_{p_T}^\prime$ in the given window.  Due to this softer constraint, the hardening of the $p_T$ spectrum of the leading jet is less dramatic than in the $b\bar b jj$ case (e.g. compare Figs.~\ref{fig:ptjcrossover} and \ref{fig:ptj_1-4j-crossover}).  The crossover of the SPS and DPS contributions occurs near $80$~GeV for $S_{p_T}^\prime < 0.2$ and near $50$~GeV for 
$S_{p_T}^\prime < 0.4$

%
%
\section{Discussion and conclusions} 
\label{sect:conclusions}

Our goal is to develop a method to search for a double parton scattering contribution in the 
$b~\bar{b}~j~j$ and 4 jet final states at LHC energies and to measure the magnitude of 
its contribution relative to the single parton contribution to the same final states.  Based 
on our parton level simulations, we find that variables such as $S_{p_T}^\prime$  and 
$S_{\phi}$ that take into account information from the entire final state, thereby including both of the hard 
subprocesses in DPS, are more effective at discrimination than variables such as $\Delta \phi_{jj}$ that 
reflect only a subset of the final-state.  The enhancement at low values of $S_{p_T}^\prime$ 
shown in Figs.~\ref{fig:sptprimecut},~\ref{fig:scatter1} and~\ref{fig:sptprimedem} provides a good signature for the presence of double parton scattering.  We urge experimenters to search for such a 
concentration of events in data at the LHC.  
Having found this enhancement, we then suggest that the magnitude of this peak be examined as 
a function of the transverse momentum $p_T^{j1}$ of the leading jet in the event sample.  
The double parton  scattering contribution in the peak region should fall off more rapidly 
with $p_T^{j1}$ than the rest of the sample.  The distribution of events in the region of small values of  $S_{p_T}^\prime$ should also be examined as a function of the inter-plane angle $\Phi$ to see whether the flat behavior is seen, as expected for two independent production processes.  Once these characteristics of double parton scattering are established, the data can be used to determine the effective normalization $\sigma_{\rm eff}$, defined and discussed  in the Introduction.  It will be interesting to see whether the values extracted for $\sigma_{\rm eff}$ are about the same in the $b~\bar{b}~j~j$ and 4 jet final states and how they compare with values measured at the Fermilab Tevatron.  

Once double parton scattering is established in data, and $\sigma_{\rm eff}$ is determined, in a relatively clean process such as $b \bar{b} j j$, double parton contributions to a wide range of other processes can be computed with more certainty about their expected rates at LHC energies.  To be sure, given the approximations described in the Introduction, some variation in the values of $\sigma_{\rm eff}$ might be expected and appropriate for different processes and in different kinematic regions.  The connection of  $\sigma_{\rm eff}$ with the effective size of the hard-scattering core of the proton may mean that $\sigma_{\rm eff}$ will have different values for $gg$, $q q$, and $q \bar{q}$ scattering.   

There are several avenues for future work.  Of great importance is the proper inclusion of next-to-leading order contributions~\cite{NLO}.   They are needed to make more robust predictions of the relative normalization of the DPS and SPS contributions, of the shape of the $p_T$ distribution of the leading jet,   and for proper softening of the sharp peaks seen near $S_{p_T}^\prime = 1$ in 
Figs.~\ref{fig:sptprimecut} and~\ref{fig:sptprimedem}, and near $S_{\phi} = \pi$ in Fig.~\ref{fig:delphi}b.

It will also be important to develop joint probabilities $H^{i,k}(x_1, x_2, \mu_A, \mu_B)$ that are more sophisticated theoretically than the first approximation represented by Eq.~(\ref{eq:approx1}) in which parton-parton correlations are absent.  A valuable development in this direction are the studies presented in Refs.~\cite{Korotkikh:2004bz, Gaunt:2009re}.  

Double parton contributions are potentially relevant for a wide range of standard model processes, many already considered in the literature~\cite{Goebel:1979mi, Paver:1982yp, Humpert:1983pw, Mekhfi:1983az, Humpert:1984ay, Ametller:1985tp, Halzen:1986ue, Mangano:1988sq, Godbole:1989ti, Drees:1996rw, Eboli:1997sv, Yuan:1997tr, Calucci:1997uw, DelFabbro:1999tf, Kulesza:1999zh, Cattaruzza:2005nu, Hussein:2006xr, Maina:2009sj, Domdey:2009bg, d'Enterria:2009hd, Akesson:1986iv,Abe:1997xk, D0:2009}, and they may also feed pertinent standard model  backgrounds to new physics processes~\cite{Sullivan:2008ki}.  They could be an issue in studies of Higgs boson production in weak-boson-fusion since the ``forward'' jets could come from a second hard subprocess.

\section{Acknowledgments} 

We benefited greatly from  discussions with Dr. Thomas LeCompte and from communications with Dr. John Campbell during the early development of this project.  We also thank Tom, John, and Professor Jianwei Qiu for valuable comments and suggestions on an earlier draft of this paper.   Research in the High Energy Physics Division at Argonne is supported by the U.~S.\ Department of Energy under Contract No.\ DE-AC02-06CH11357.  The research of GS at Northwestern is supported by 
the U.~S.\ Department of Energy under Contract No.\ DE-FG02-91ER40684.


%
%

\end{document}